\documentclass[manuscript]{aastex}
\usepackage{aas_macros,url,natbib,amssymb,amsmath,CJK}

\shorttitle{Where are the Mini Kreutz-family Comets?}
\shortauthors{Ye et al.}

\begin{document}
\begin{CJK*}{UTF8}{gbsn}

\title{Where are the Mini Kreutz-family Comets?}

\author{Quan-Zhi Ye (叶泉志)}
\affil{Department of Physics and Astronomy, The University of Western Ontario, London, Ontario N6A 3K7, Canada}
\email{qye22@uwo.ca}

\author{Man-To Hui (许文韬)}
\affil{Department of Earth, Planetary and Space Sciences, University of California at Los Angeles, 595 Charles Young Drive East, Los Angeles, CA 90095-1567, U.S.A.}

\author{Rainer Kracht}
\affil{Ostlandring 53, D-25335 Elmshorn, Schleswig-Holstein, Germany}

\and

\author{Paul A. Wiegert}
\affil{Department of Physics and Astronomy, The University of Western Ontario, London, Ontario N6A 3K7, Canada}

\begin{abstract}
The Kreutz family of sungrazing comets contains over 2~000 known members, many of which are believed to be under $\sim 100$~m sizes (\textit{mini~comets}) and have only been studied at small heliocentric distances ($r_\mathrm{H}$) with space-based SOHO/STEREO spacecraft. To understand the brightening process of mini Kreutz comets, we conducted a survey using CFHT/MegaCam at moderate $r_\mathrm{H}$ guided by SOHO/STEREO observations. We identify two comets that should be in our search area but are not detected, indicating that the comets have either followed a steeper brightening rate within the previously-reported rapid brighten stage (the \textit{brightening~burst}), or the brightening burst starts earlier than expected. We present a composite analysis of pre-perihelion light-curves of five Kreutz comets that covered to $\sim 1$~AU. We observe a significant diversity in the light-curves that can be used to grossly classify them into two types: C/Ikeya-Seki and C/SWAN follow the canonical $r_\mathrm{H}^{-4}$ while the others follow $r_\mathrm{H}^{-7}$. In particular, C/SWAN seems to have undergone an outburst ($\Delta m>5$~mag) or a rapid brightening ($n\gtrsim11$) between $r_\mathrm{H}=1.06$~AU to $0.52$~AU, and shows hints of structural/compositional differences compared to other bright Kreutz comets. We also find evidence that the Kreutz comets as a population lose their mass less efficiently than the dynamically new comet, C/ISON, and are relatively devoid of species that drive C/ISON's activity at large $r_\mathrm{H}$. Concurrent observations of C/STEREO in different wavelengths also suggest that a blue-ward species such as CN may be the main driver for brightening burst instead of previously-thought sodium.
\end{abstract}


\section{Introduction}

First identified by \citet{1888QB726.82.K92...}, the Kreutz family of sungrazing comets is probably the most well-known comet family, mostly because of its small perihelion distance (almost at the surface of the Sun) and the large number of known members ($\sim 2~000$ as of the time of writing). Some of the most spectacular comets in history also belong to the Kreutz family, such as the Great Comet of 1843 (C/1843 D1 or 1843 I), the Great Comet of 1882 (C/1882 R1 or 1882 II), and Comet Ikeya-Seki in 1965 (C/1965 S1 or 1965f). However, until the very recent decades, the investigation of the Kreutz family had been restricted to the few brightest members, mostly because of the very unfavorable visibility for ground-based observers due to the extreme orbit geometry; most members are too faint and too close to the Sun to be observed from the ground.

Space-based solar coronagraphs launched in the 1970s and 1980s have proved to be good comet observatories as well \citep[e.g.][and many others]{1991Icar...90...96M}, as they can detect sungrazing comets very close to the sun. The Solar and Heliospheric Observatory (SOHO), launched into a ``halo orbit'' near Earth's L1 point in 1995 \citep{1995SoPh..162....1D}, is still operational and has found $\sim2~800$ comets as of September 2014, most of which are Kreutz family members with sizes believed to be under $\sim100$~m (for convenience, we term such objects \textit{mini~comets} without any physical implication). Mini comets are essentially transitional objects from comets to meteoroids, they are direct products in cometary fragmentation and form the base of the Kreutz hierarchy. While the success of SOHO has allowed us to look into some general questions about the Kreutz family such as the dynamical evolution of the hierarchy \citep[e.g.][]{2004ApJ...607..620S,2007ApJ...663..657S,2013ApJ...778...24S}, further studies are limited by short observation arcs (within a few days of the perihelia), low astrometric precision and relatively high image noise of SOHO.

Ground-based observation of mini Kreutz comets, on the other hand, remains difficult and no successful observations have been reported. The only reported attempt was made by \citet[][hereafter K10]{2010AJ....139..926K}, who used the 4-m KPNO telescope and the 6.5-m Magellan telescope at Las Campanas to search for Kreutz comets 3--6 months prior to their expected arrivals at perihelia, with no positive detection reported. Robert McNaught (personal comm. 2007) also conducted similar searches at comparable elongations, with the 0.5-m Uppsala telescope at Siding Spring Observatory, but also did not find any comet.

One of the top questions involves the brightening process of these comets, which may bear implications about the structure and composition of the comets \citep[e.g.][]{2002Icar..159..529K}. K10's analysis of some 1~000 Kreutz comets observed by SOHO shows a stage of rapid brightening following $n\sim7$ from an unknown heliocentric distance $r_\mathrm{H}$ (K10 noted that it is ``unlikely'' to be beyond $50~\mathrm{R_\odot}$) to $\sim24~\mathrm{R_\odot}$, where $n$ is defined such that the comet brightens following $r_\mathrm{H}^{-n}$ where $r_\mathrm{H}$ is the heliocentric distance. For convenience of discussion, we call this feature the \textit{brightening burst}. It is unclear how far the brightening burst extends, but understanding this behavior would give clues about the volatile content and compositional diversity within the Kreutz family.

In this paper, we conducted a search for mini Kreutz comets at moderate heliocentric distance ($\sim 1$~AU) to explore the brightening behavior of these comets. This was done by surveying the ``sweet spots'' defined through the Kreutz Population Model (\S2 and \S3), appropriate to comets $\sim1$~AU from the Sun. The results are then combined with the observations from space-based coronagraphs (such as SOHO) to produce composite light-curves (\S4). The results are discussed and compared to several bright historical Kreutz comets observed at moderate $r_\mathrm{H}$ (\S5).

\section{Methodology}

The century-old method of comet searching is to scan the regions that are statistically most likely to contain comets. This does not require \textit{a priori} knowledge of the position of individual comets. However, for the case of Kreutz comets, there are two issues: (i) the dynamical evolution of the Kreutz family is quite complicated and many details are not clear \citep[e.g.][]{1967AJ.....72.1170M,1989AJ.....98.2306M,2004ApJ...607..620S,2013ApJ...778...24S}, and (ii) the orbital elements of almost all currently known Kreutz comets are derived from low-resolution SOHO observations, which results in large uncertainties. The two issues cause large uncertainties of the on-sky positions of the Kreutz comets at moderate $r_\mathrm{H}$, on the order of several tens of square degrees at $1$~AU.

Additionally, we can now make use of the new Solar TErrestrial RElations Observatory (STEREO) spacecraft that provides rather a long baseline when coupled with SOHO observations to help comet orbit determination (Table~\ref{tbl-spacecrafts}). STEREO consists of two identical satellites that orbit the Sun on about the same orbit of the Earth, but one is leading the Earth (STEREO-A) and the other is trailing the Earth (STEREO-B). The most useful instrument for improving the orbit determination is perhaps COR-2, an instrument on-board STEREO that comes with high resolution and a large field-of-view, but is insensitive to faint comets due to its observing bandpass.

For the case of searching for mini Kreutz comets, one needs to cover a large area (at the order of 10 square degrees) with deep limiting magnitude ($\gtrsim 20$~mag). Since earlier studies have suggested that Kreutz comets may have a steep brightening rate that makes detection far beyond Earth's orbit exceedingly difficult, one also needs to search at the smallest elongation possible. Lastly, the search zones for Kreutz comets are located in the southern skies, which severely limits the use of northerly facilities.

\section{Kreutz Population Model}

\begin{figure}
\includegraphics[width=0.5\textwidth]{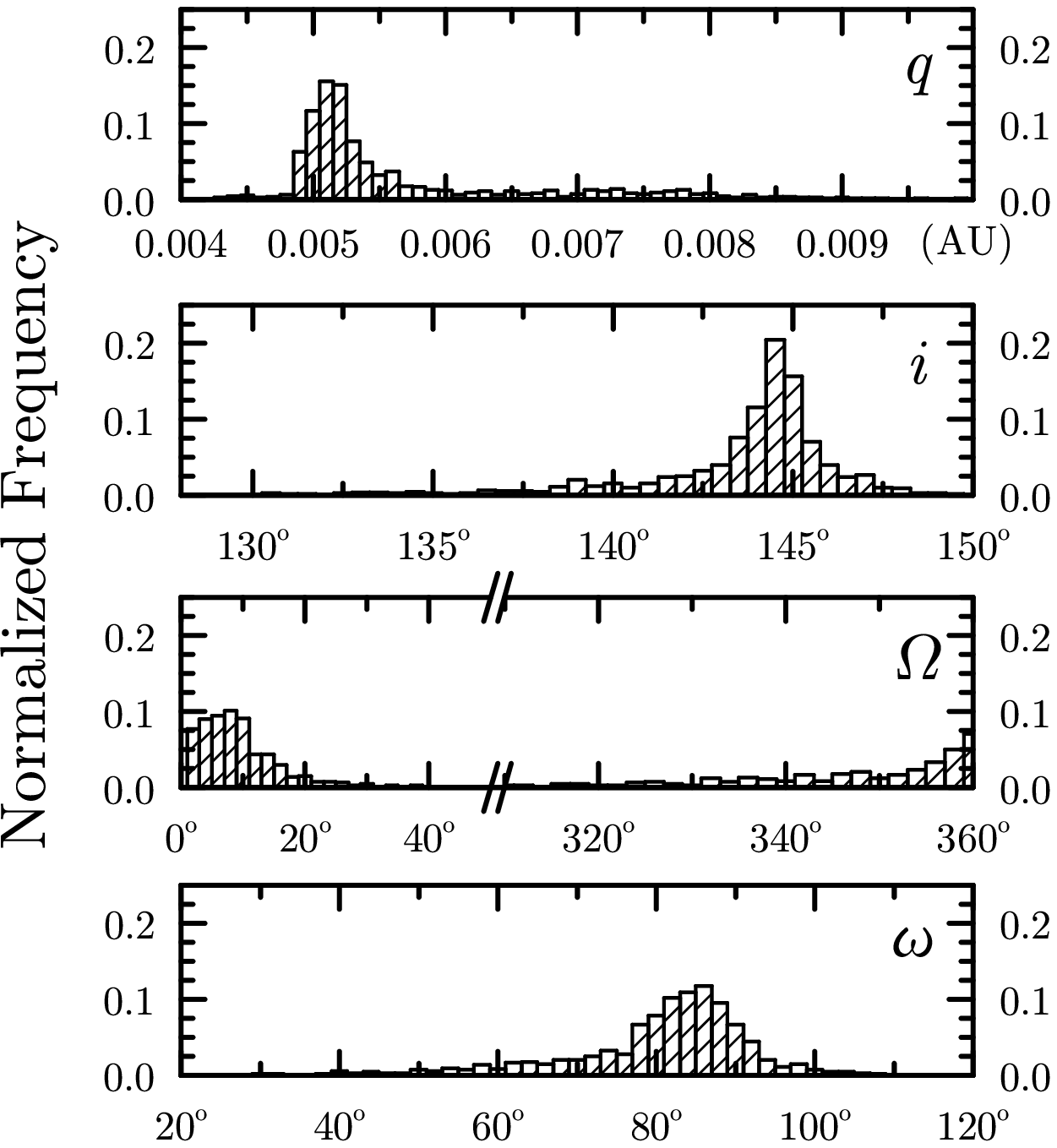}
\caption{Distribution of orbital elements (perihelion distance $q$, inclination $i$, longitude of the ascending node $\Omega$, and argument of perihelion $\omega$) from 1~283 Kreutz comets selected from JPL small-body database.}
\label{fig-krz-orb}
\end{figure}

To determine our surveying areas, we consider the distribution of orbital elements of 1~253 known Kreutz comets, most of which were discovered in 1996--2008 by SOHO (Figure~\ref{fig-krz-orb})\footnote{Many post-2008 comets do not have published orbits at the time of study.}. The orbital elements are obtained from the JPL small-body database\footnote{\url{http://ssd.jpl.nasa.gov/sbdb_query.cgi}, retrieved 2012 Aug. 17.}. According to this distribution of orbital elements, we then generate 10~000 pseudo comets $30\pm15$~days ahead of their perihelion passages and centered at our desired observing time. We also assign an absolute magnitude for each pseudo comet following K10's size distribution, assuming geometric albedo $A_p=0.05$, which allows us to calculate the apparent magnitude of the comet using K10's ``worst case'' scenario: comets follow $n=4$ until $50~\mathrm{R_\odot}$, then follow $n=7$ until $24~\mathrm{R_\odot}$, where they move back $n=4$ and remain so until reaching the average distance of peak brightness at $12~\mathrm{R_\odot}$. We find the median brightness at $\Delta=r_\mathrm{H}=1~$AU ($\Delta$ is the geocentric distance) is about 21~mag. Depending on the season, the median motion of the pseudo comets varies from a few arc seconds per minute to 10--20 arc seconds per minute.

We recognize the weakness of randomly distributing each orbital element as this assumes that each element is independent from each other. However, it has been known that the Kreutz family is composed by at least two (probably more) sub-groups, and the abundance and spread of each sub-group is not clearly known. We think that the current model is practically the best approach at the moment, which, when coupled with a ``worst scenario'' brightening model, should give us an idea about the lower limits of on-sky comet densities.

\section{Observation and Analysis}

\subsection{CFHT Observation and Initial Analysis}

\begin{figure}
\includegraphics[width=0.5\textwidth]{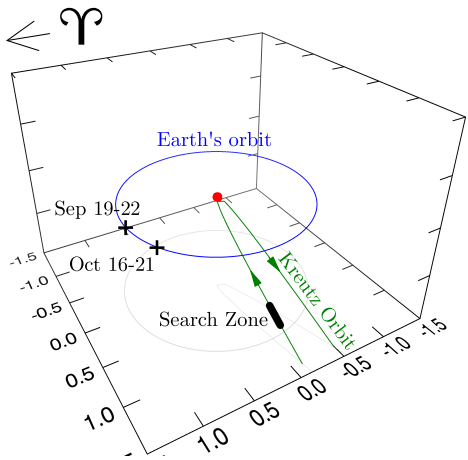}
\caption{Observation circumstances and the search zone showed in a heliocentric frame. The arrow at the top-left corner points to the Vernal Equinox ($\Upsilon$). The two plus signs indicate the position of the Earth at the time of observation. The unit of the axes are in AU. The cometary orbital plane is below the ecliptic plane. The size of the Sun is not to scale.}
\label{fig-obs-setup}
\end{figure}

We used the 3.6-m Canada-France-Hawaii Telescope (CFHT) with the 1 square degree MegaCam for our survey. A northerly latitude ($19.8^{\circ}$~N) prohibits observation when the Earth is closest to the common orbit of the Kreutz comets (which occurs in late December and solely favors southern hemisphere observers), but observation is still feasible in September and October, when the Earth is at moderate distance ($\sim 1~$AU) to the inbound leg of the Kreutz comets (Figure~\ref{fig-obs-setup}). At $25^{\circ}$ elevation and with a seeing of 1.5'', we expected to reach $m_\mathrm{g'}$=23~mag with 30~s exposures. Although bandpasses at longer wavelengths such as r' are slightly less responsive to atmospheric extinction, we chose g' as it includes most of the optical emission species (mostly C$_2$ and C$_3$ species) that may increase the chance of detection. We took three 30~s exposures at each position, separated by $\sim 5$~min to distinguish moving objects. To maximize the overall coverage, we did not make attempts to fill the gaps between individual CCDs, hence the filling factor at each pointing location is $93\%$ \citep{2000SPIE.4008..427D}.

\begin{figure}
\includegraphics[width=0.5\textwidth]{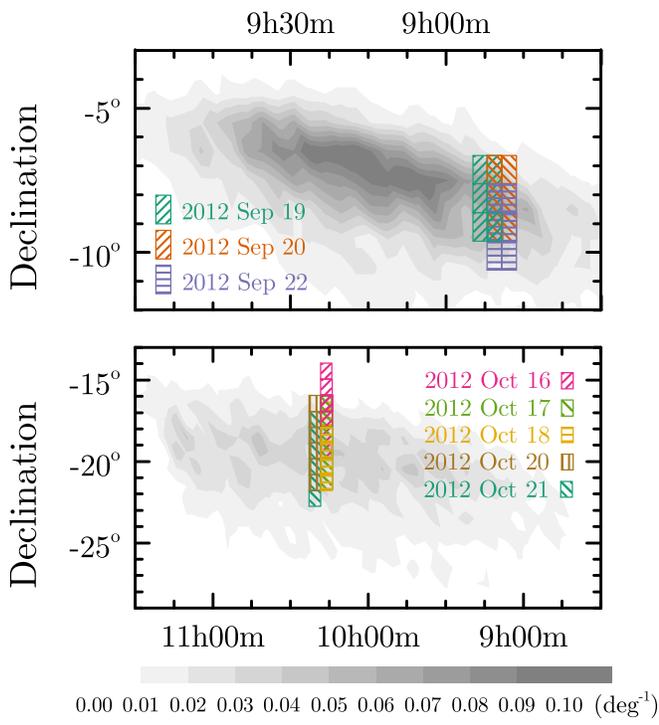}
\caption{Simulated density of comets on 2012 Sep. 20 and Oct. 20, as well as the observed areas}
\label{fig-obs-area}
\end{figure}

We ran the Kreutz Population Model targeting for 2012 Sep. 20 and 2012 Oct. 20 for pseudo-comets that would reach their perihelia between 2012 Oct. 5 to 2012 Dec. 5. Considering the arrival rate of comets visible to C3 (i.e. peak brightness $<8$~mag) in this season is $\sim 10$ per month as noted by K10, we find a maximum density of 0.04--0.08 comets per square degree. We note that the actual density in the ``sweet spots'' should be higher if the orbital elements are dependent on each other. The observations were executed around the two dates, as summarized in Table~\ref{tbl-obs}. The observed areas are shown in Figure~\ref{fig-obs-area}.

After standard calibration (i.e. bias/dark field subtraction and flat field division), the images are processed by a series of semi-automatic routines that looked for moving sources as developed by \citet{2007AJ....133.1609W,2009Icar..201..714G,2010Icar..210..998G} and \citet{2013AJ....145..152A}. Automatic detections are manually verified to remove false positives. No possible Kreutz comets were found in the data.

To determine the detection efficiency, we seed artificial sources that mimic brightness and movement of possible Kreutz comets into one arbitrarily chosen data-set from each night. The seeded images are processed by the same routines used for real data. The magnitudes of detected artificial sources are binned into 0.5~mag to determine completeness of detection along magnitude ranges. The lower boundaries of magnitude bins above $1\sigma$ detection efficiency are reported in Table~\ref{tbl-obs}.

\subsection{Combined Analysis with SOHO/STEREO Data}

After the expected perihelion passages of possible Kreutz comets that should be in our images, we check comets detected by SOHO/STEREO to see if there is any comet with an orbit accurate enough to pinpoint its position in our images. SOHO found 36 Kreutz comets in the time window set up for simulation (2012 Oct. 5 to 2012 Dec. 5), but due to the large on-sky uncertainties of SOHO-based Kreutz comets, we focus on the comets that are concurrently detected in COR-2. We identify 5 such comets (Table~\ref{tbl-adhoc}), but none of these five comets are published\footnote{One comet, temporarily sequenced as SOHO-2387, was published later using our reported measurements and received an official designation as C/2012 U3 (STEREO) (the comet was first detected in SOHO data despite being named after STEREO). We will stick to the official designation C/2012 U3 (STEREO) in this study, but use unofficial SOHO sequence numbers for unpublished comets.}, therefore we need to compute their orbits first. We obtain the raw ``level-0.5'' data from the STEREO public database, and calibrate them using the SolarSoft IDL package \citep{1998SoPh..182..497F}. We then perform astrometric measurements using the NOMAD catalog \citep{2004AAS...205.4815Z} and calculate their orbits using the EXORB program\footnote{\url{http://chemistry.unina.it/~alvitagl/solex/Exorb.html}, retrieved 2014 July 21.} (Table~\ref{tbl-adhoc-orb}). The uncertainties of the orbital elements are computed using a Monte Carlo subroutine within EXORB, which allow us to estimate the on-sky uncertainty of the comet.

\begin{figure*}
\includegraphics[width=\textwidth]{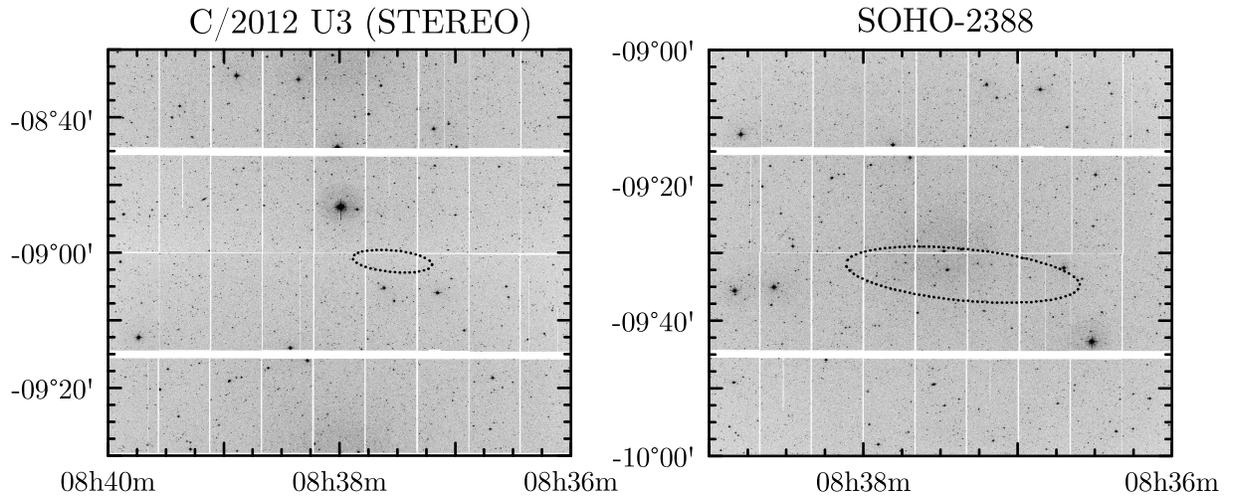}
\caption{On-sky uncertainties of C/2012 U3 (STEREO) (left) and SOHO-2388 (right) on real CFHT observations on 2012 Sep. 20 (C/STEREO) and Sep. 22 (SOHO-2388).}
\label{fig-area}
\end{figure*}

Two comets, C/STEREO and the unpublished SOHO-2388, are found to be within the field-of-view of images taken in 2012 Sep. 20 and 22 (Figure~\ref{fig-area}). We mask out background stars and stack the images at the calculated motions of the comets to pull the limits, but still do not find any comets. The detection limits (assuming SNR=3) of the stacked images are $m_\mathrm{g'}$=23.7 for C/STEREO and $m_\mathrm{g'}$=24.0 for SOHO-2388. We convert the magnitudes to Johnson V magnitude using the using CFHT transformation equations\footnote{\url{http://www2.cadc-ccda.hia-iha.nrc-cnrc.gc.ca/en/megapipe/docs/filt.html}, retrieved 2014 July 21.} and \citet{2006A&A...460..339J} assuming solar colors, which gives $m_\mathrm{V}$=22.3 and $m_\mathrm{V}$=22.5 normalized to $\Delta=r_\mathrm{H}=1~$AU.

We also measure the light-curves of C/STEREO and SOHO-2388 in SOHO/STEREO data. The reduction procedure follows K10 for SOHO data and \citet{2013MNRAS.436.1564H} for STEREO data, including normalization to $\Delta=1$~AU and the phase angle $\theta=90^{\circ}$ following the compound Henyey-Greenstein model \citep{2007ICQ....29...39M}:

\begin{equation}
 \phi(\theta) = \frac{\delta_{90}}{1+\delta_{90}} \left\{ k \left[ \frac{1+g_f^2}{1+g_f^2-2g_f \cos{(\pi-\theta)}} \right]^{\frac{3}{2}} + (1-k) \left[ \frac{1+g_b^2}{1+g_b^2 - 2 g_b \cos{(\pi-\theta)}} \right]^{\frac{3}{2}} + \frac{1}{\delta_{90}} \right\}
\end{equation}

where $\phi(\theta)$ is the scattering function which relates to magnitude by $m(\theta)=2.5 \log{\phi(\theta)}$, $\delta_{90}$ is the dust-to-gas flux ratio at $\theta=90^{\circ}$, $k$ is the partitioning coefficient between forward and back-scattering, and $g_f$ and $g_b$ are forward and back-scattering asymmetry factors. \citet{2007ICQ....29...39M} found $k=0.95$, $g_f=0.9$ and $g_b=-0.6$ by fitting the data to observations of six comets. For optical observations, $\delta_{90}=1$ is used for normal and $\delta_{90}=10$ is used for dusty comets. However, K10 suggests Kreutz comets have stronger sodium emission compared to other comets and recommends using $\delta_{90}=0.52$ for SOHO's broad-band observations. For COR-2 and HI-1, as no $\delta_{90}$ have been reported, we tentatively assign $\delta_{90}=10$ for COR-2 and $\delta_{90}=1$ for HI-1 given that no emission lines have been found in the main transmission region of COR-2 and HI-1 using the spectrum of C/Ikeya-Seki \citep[i.e. 630--750~nm, see][
and many others]{1967ApJ...147..718P}. The $\delta_{90}$ value for HI-1 is smaller as HI-1 has the ``blue leak'' issue \citep{2010SoPh..264..433B} that makes the flux contaminated by some gaseous species such as CN.

\begin{figure*}
\includegraphics[width=\textwidth]{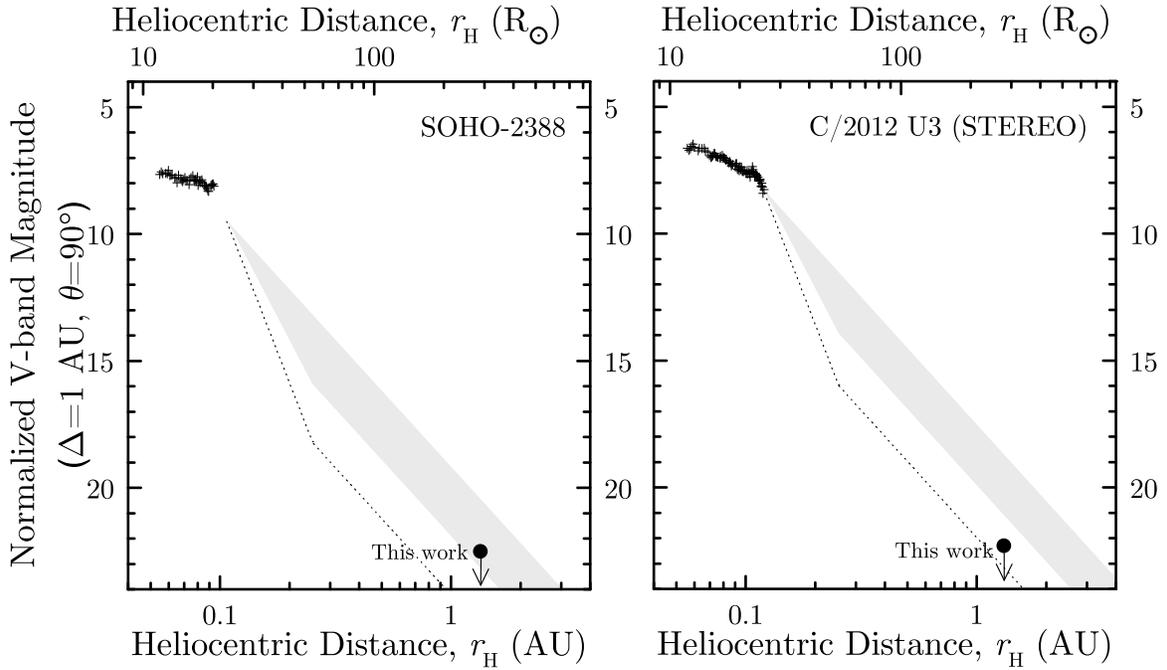}
\caption{Light-curve of C/2012 U3 (STEREO) (left) and SOHO-2388 (right) using C3 data, as well as the upper limits constrained by CFHT observations. Shaded area is the set of possible light curves from no brightening burst ($n=4$) to nominal rate of brightening burst ($n=7$) to $50~\mathrm{R_\odot}$ as suggested by K10. Dotted line is steepest rate of brightening burst ($n=9$) suggested by K10.}
\label{fig-lc-2387-8}
\end{figure*}

Ground-based observations are only compared with C3 observations (Figure~\ref{fig-lc-2387-8}), as observations from other instruments suffer from small coverage (C2 and COR-2), or blue leak contamination (HI-1).

We also note that only a subset of COR-2 data is used in this work, as all COR-2 images are polarized. Here we only use COR-2 images taken with two polarizers oriented to $0^{\circ}$ and $90^{\circ}$, that can be approximated to ``total brightness'' images, due to the fact that cometary light is dominantly linearly polarized. Analysis of the polarized data will be reported in a separate paper.

\subsection{A Note of October 2012 Observation}

No Kreutz comets with good orbits match our October observations. Our simulation shows that any comet that may position in our October images would arrive its perihelion in 2012 Nov. 8--15. The SOHO discovery catalog\footnote{\url{https://www.ast.cam.ac.uk/~jds/knos12.htm}, retrieved 2014 Jul. 21.} shows 7 comets (SOHO-2389 to SOHO-2395) that reach perihelion within this time window and could have been positioned in the October images. Considering that about $95\%$ of bright SOHO-observed comets move in orbits close to C/1843 D1 \citep[][]{2013ApJ...778...24S}, the projected on-sky area of which is well covered in our survey, it is likely that all these 7 comets ($7\times95\%=6.65\approx7$) are within our images from a statistical perspective. Of these 7 comets, 3 are discovered in C3 and therefore should be bright enough for our survey should they follow K10's nominal rate. Since no comets are detected, we conclude that these comets, if they were indeed within our images, were no brighter than 21.7 mag (normalized to $\Delta=1$~AU).

\subsection{Composite Light-curves of the Kreutz Family}

\begin{figure*}
\includegraphics[width=\textwidth]{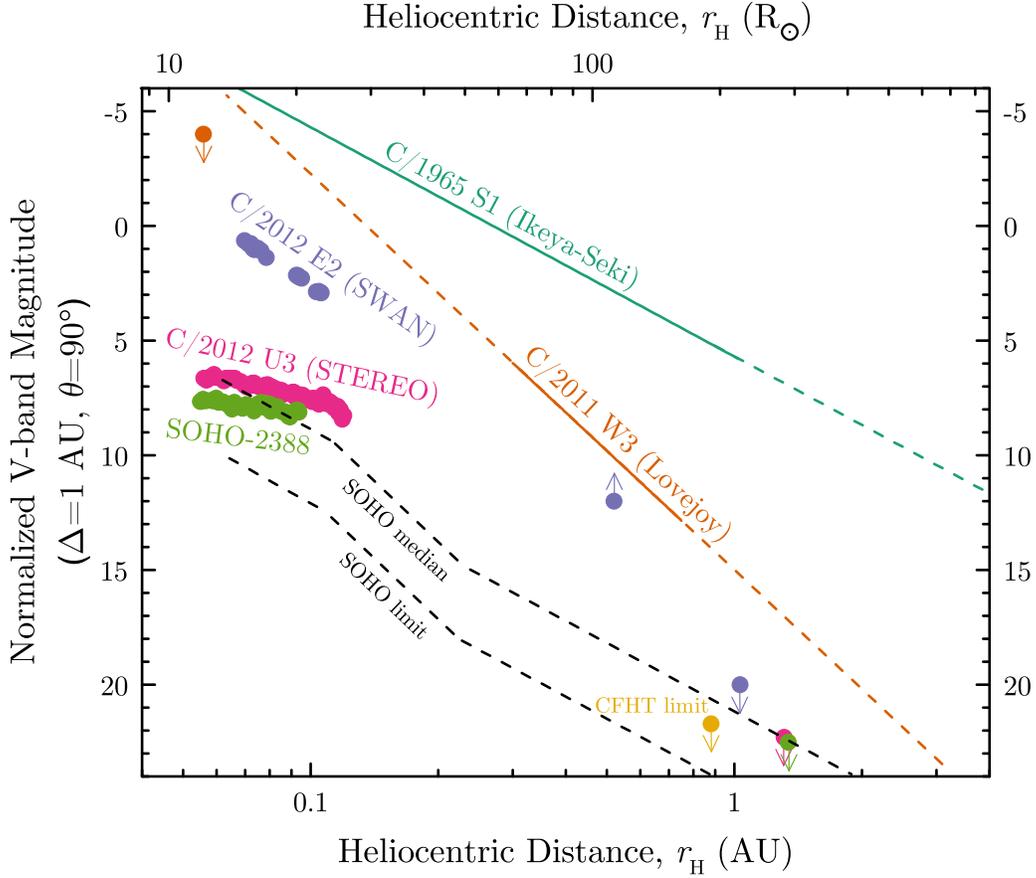}
\caption{Composite light-curves of Kreutz comets with observation up to $\sim 1$~AU. The brightening rate of C/1965 S1 (Ikeya-Seki) is reported by \citet{2002ApJ...566..577S}. The brightening rate of C/2011 W3 (Lovejoy) is reported by \citet{2012ApJ...757..127S}. Solid lines are observation-based bright-curves, while dashed lines are extrapolations following the brightening rates. The light-curves of C/2012 E2 (SWAN), C/2012 U3 (STEREO) and SOHO-2388 are from this work. The theoretical light-curves for median ($50\%$ percentile) and faintest Kreutz comets from K10's sample is also plotted. The brightening burst assumes a brightening rate of $n=7$.}
\label{fig-lc}
\end{figure*}

To produce a composite set of pre-perihelion light-curves of Kreutz comets, we measure two other bright Kreutz comets that emerged in recent years and that have been observed or constrained by ground-based telescopes, C/2011 W3 (Lovejoy)\footnote{We measure C/Lovejoy in HI-1 data until it reaches $\sim1.5$~mag and saturate the detector. No COR-2 measurement can be made as the images are saturated.} and C/2012 E2 (SWAN)\footnote{Only HI-1B and COR-2B measurements can be made. The comet was not visible in HI-1A, and COR-2A misses data taken with combined polarizers.}, and compile our results with the results from previous studies. Ground-based observations and C3 measurements are shown as Figure~\ref{fig-lc}. STEREO measurements are conducted following the same procedure used on C/STEREO and SOHO-2388. They are then converted to Johnson-Cousins system following the method proposed by \citet{2010SoPh..264..433B} and modified by \citet{2013MNRAS.436.1564H}. The light-curve fits of C/1965 S1 (Ikeya-Seki) and C/2011 W3 (Lovejoy) are taken from \citet{2002ApJ...566..577S} and \citet{2012ApJ...757..127S}. The upper limit of peak brightness of C/Lovejoy is reported by Karl Battams\footnote{\url{http://sungrazer.nrl.navy.mil/index.php?p=news/birthday_comet}, retrieved 2014 Jul. 30.}.

We pay special attention to C/SWAN, the only Kreutz comet detected by SWAN until now. SWAN, or the Solar Wind ANisotropies camera, is an all-sky camera on-board SOHO spacecraft that makes daily observation of Lyman-$\alpha$ emission. SWAN's low spatial resolution (at the order of $1^{\circ}$) makes photometric calibration difficult, therefore we use the detection limit ($\sim12$~mag) constrained using ``gassy'' comet, 2P/Encke, as the lower limit of C/SWAN in SWAN observations (a dustier comet needs to be brighter to be visible in SWAN image, potentially resulting an overestimation of the lower limit of C/SWAN's brightness). We also locate four Siding Spring Survey (SSS) images \citep[c.f.][]{2009ApJ...696..870D} taken on 2012 Feb. 14 that cover the predicted on-sky uncertainty area of C/SWAN (of which $r_\mathrm{H}=1.06~$AU). We mask out the background stars and stack the images at the predicted motion rate of the comet, which does not show any comet down to $m_\mathrm{V} = 20.0$~mag. We also check other bright Kreutz comets in SSS's exposure catalog, including C/2009 Y4, C/2010 B3, C/2010 E6, C/2010 V8 and C/2010 W2, but there are no matches.

\section{Discussion}

\subsection{C/1965 S1 (Ikeya-Seki) and C/2011 W3 (Lovejoy)}

C/Ikeya-Seki brightened steadily from discovery ($r_\mathrm{H}=1$~AU) to $\sim 0.03$~AU with $n=4$. The comet was not observed from $50-20~\mathrm{R_\odot}$, but the lack of offset between the two segments beside the unobserved segment of the light-curve supports the idea that the comet did not deviate much from $n=4$ in the unobserved segment.

C/Lovejoy also brightened steadily for the most part since discovery ($r_\mathrm{H}=0.75$~AU) but followed $n=6.9$ \citep{2012ApJ...757..127S}. HI-1 observation indicates that the brightening rate gradually decreased from $n=9.8$ to $n=4.1$ from $r_\mathrm{H}=0.3$ to $0.2$~AU. It is interesting to note that the brightening rate near $r_\mathrm{H}=0.3$~AU constrained by HI-1 is steeper than the one derived from the ground. Alternating $\delta_{90}$ would not help much to remove this discrepancy, since the observations are conducted at $\theta \sim 85^{\circ}$, thus bear minimal phase angle effects. The feature could be explained by short-term fluctuations of the comet's brightness, as hinted by the data points in \citet[][Figure 6]{2012ApJ...757..127S}. Another possible cause is the relative increase of species HI-1 detectors sensitive to but not to most ground-based observers, prominently blue-ward species like CN (388.3~nm). Simultaneous broad-band measurements would help to confirm or rule out this possibility, but unfortunately no measurements can be taken from C3 as the comet is saturated.

The fact that C/Lovejoy follows $n=6.9$ within $r_\mathrm{H}=0.75$ to $0.25$~AU ($161$ to $54~\mathrm{R_\odot}$) seems to argue against K10's idea that the brightening burst is unlikely to start beyond $50~\mathrm{R_\odot}$. Since the comet must follow $n \sim 2$ when it is far away from the Sun and behaves like an asteroid, a steep brightening rate like $n=6.9$ must begin somewhere during the journey, presumably controlled by the dominant volatile when the comet becomes active. This would indicate that C/Lovejoy has an abundant supply of such volatiles so that it can maintain the steeper $n=6.9$ until $r_\mathrm{H}=0.25$~AU. The gradual decrease to $n=4$ also seems to indicate that the volatile is being depleted. An alternative but also plausible scenario is that C/Lovejoy underwent a series of minor outbursts that helped it to maintain a steeper brightening rate. Extending the light-curve to a larger $r_\mathrm{H}$ would be useful to test both hypotheses, but an examination of SSS's exposure catalog does not turn up any pre-discovery images.

No useful light-curve can be obtained shortly before C/Lovejoy's perihelion (within $r_\mathrm{H}=0.1$~AU) due to saturation, but we conclude that C/Lovejoy follows $n=4$ to its peak brightness, as this agrees the upper limit of the peak magnitude.

\subsection{C/2012 E2 (SWAN)}

C/SWAN is truly an exception as no other known SOHO-observed Kreutz comet -- including those with peak magnitudes comparable to or brighter than C/SWAN -- has been bright enough to be seen by SWAN.

SSS's negative detection can be used to constrain the upper limit of the cometary nucleus. Assuming $A_p=0.05$, the diameter of the nucleus can be estimated by $D=5943\times10^{-\frac{H_0}{5}}$ with $D$ in km and $H_0$ is the absolute magnitude of the comet. We get $D=680$~m as the upper limit for C/SWAN, appropriate to $r_\mathrm{H}=1.06$~AU. The non-detection also suggests that C/SWAN experienced an outburst with $n\gtrsim11$ between $r_\mathrm{H}=1.06$~AU to $0.52$~AU, then steadily followed $n=4$ from 0.52~AU until perihelion as suggested by SWAN, C3 and HI-1 light-curves. Following K10's methodology, we estimate C/SWAN was $\sim80$~m in diameter at its peak brightness.

\begin{figure*}
\includegraphics[width=\textwidth]{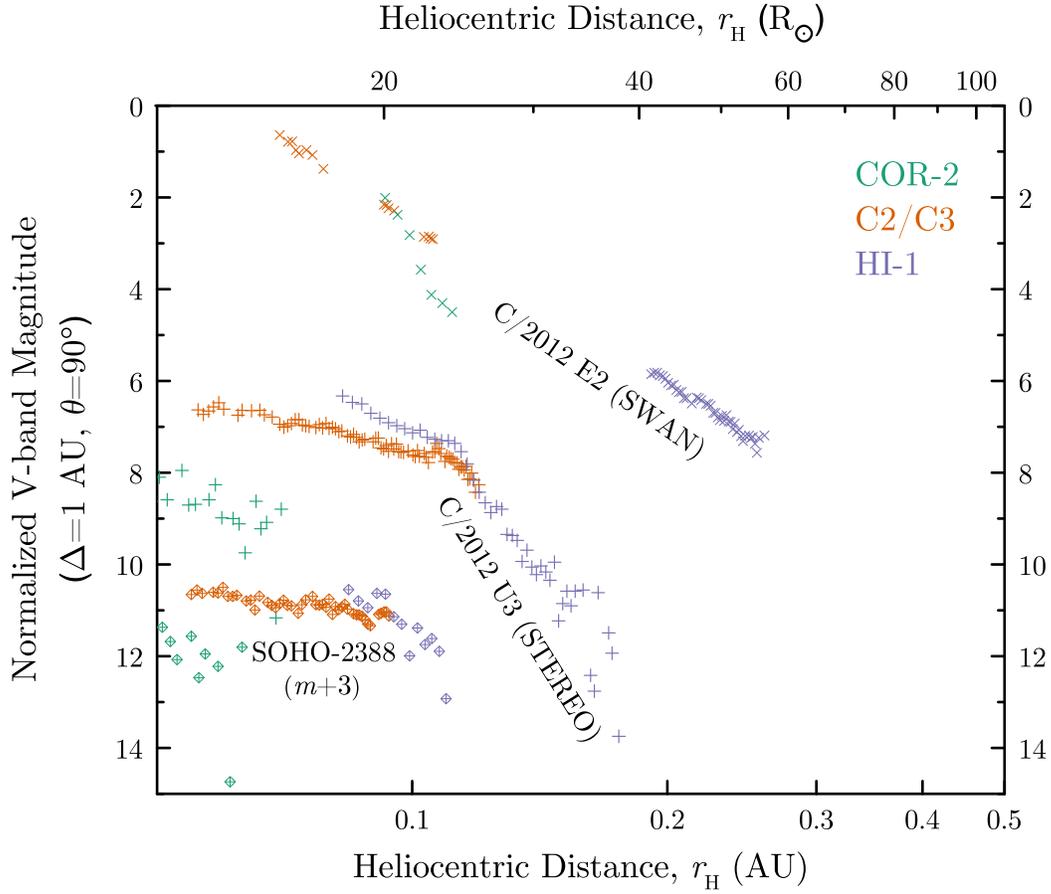}
\caption{Light-curves of C/2012 E2 (SWAN), C/2012 U3 (STEREO) and SOHO-2388 (offset by 3 magnitude for clarity) in COR-2, C2/C3 and HI-1. Different symbols represent different comets, from top to bottom or right to left: C/SWAN, C/STEREO and SOHO-2388.}
\label{fig-comet-bandpass}
\end{figure*}

While the light-curves of SWAN, C3 and HI-1 fit well into each other, the COR-2 light-curve shows a short, rapid brightening just beyond $20~\mathrm{R_\odot}$ (Figure~\ref{fig-comet-bandpass}). This is puzzling as such a feature is absent in the C3 data (some C3 data are too noisy to calibrate and are therefore ignored, leaving gaps in the C3 light-curve). The major cometary species in the COR-2 bandpass would be $\mathrm{NH_2}$, but previous studies of C/Ikeya-Seki \citep[e.g.][and many others]{1967ApJ...147..718P} showed no $\mathrm{NH_2}$ emissions at $r_\mathrm{H}=0.1$~AU and beyond. It is unlikely that C/SWAN is different than C/Ikeya-Seki in composition and possesses some $\mathrm{NH_2}$, as they are descendants of a common parent and should be similar in composition. A more likely explanation is that, since gas emission is primarily in broad-band wavelengths (C3) but not in narrow-band wavelengths (COR-2), a sudden decrease in gas emission would result a ``shallower'' increase in broad-band magnitude comparing to narrow-band magnitude. This probably indicates the erosion of the less volatile portion of the nucleus.

Since the transmission ranges of COR-2 and HI-1 largely overlap, the magnitude difference between these two instruments is worth some attention. The main transmission bandpasses for COR-2 and HI-1 are 650--750~nm and 630--730~nm respectively, which should bear little difference as no known emission species for Kreutz comets falls in this region. The most plausible explanation is the gas contamination in the ``blue leak'' region, most likely CN (388.3~nm). Deriving a number for the production rate is possible but risky due to the detector's uneven response across different wavelengths. On the other hand, Fe I emission (dominantly 371.9~nm and 404.5~nm) has also been reported in C/Ikeya-Seki's post-perihelion spectrum \citep[e.g.][]{1967ApJ...147..718P}, but we doubt that Fe I can be a major contribution as the comet must be cooler at the same $r_\mathrm{H}$ in its pre-perihelion leg.

\begin{figure}
\includegraphics[width=0.5\textwidth]{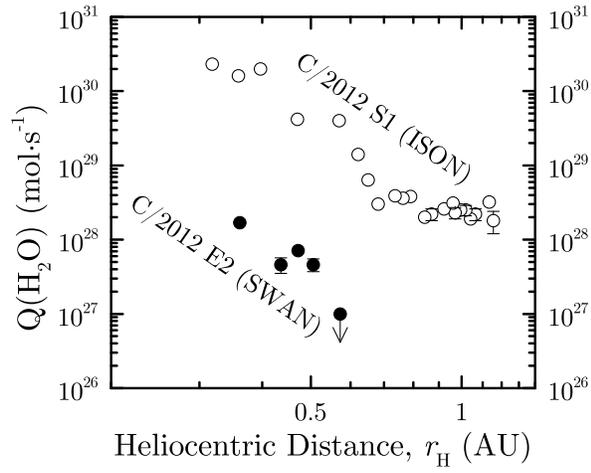}
\caption{Water production rates of C/2012 E2 (SWAN) and C/2012 S1 (ISON) as derived by \citet{2012IAUC.9252....2C,2013IAUC.9266....1C}. Some error bars are too small to appear on the figure.}
\label{fig-swan}
\end{figure}

We also examine the water production rate of C/SWAN derived from SWAN data \citep{2013IAUC.9266....1C}, shown in Figure~\ref{fig-swan}. The water production rate of C/2012 S1 (ISON) \citep{2013IAUC.9266....1C}, a dynamically new sungrazing comet that does not belong to the Kreutz family, is also drawn for comparison. For C/SWAN, the increase of water production rate follows $n=5.4$ from $r_\mathrm{H}=0.57$ to $0.36$~AU after the presumed outburst. It is noteworthy that C/ISON follows a similar slope ($n=5.3$) after an outburst as noted by ground-based observers \citep[e.g.][]{2013CBET.3718....2C}, although this could well be coincidence, as we find no reason to believe that the composition and structure of the two comets is similar.

The compositional difference between C/SWAN (and its fellow Kreutz comets) and C/ISON is further suggested by the non-detection of any mini Kreutz comets by ground-based surveys. C/ISON was evidently more active at large $r_\mathrm{H}$, being discovered at $m=19$ at $r_\mathrm{H}=6$~AU. At the peak distance of mini Kreutz comets, C/ISON was $\sim 10$~mag brighter than the faintest Kreutz comet detectable by C3. Considering that C/ISON had reached $m=10$~mag at $r_\mathrm{H}=1.5$~AU, if C3-detectable mini Kreutz comets are as active as C/ISON, they should have reached the detection limits of modern Near-Earth Object surveys such as the SSS ($\sim 20$~mag) at similar $r_\mathrm{H}$. The fact that no Kreutz comets at these sizes have been detected may suggest that the surface layers of mini Kreutz comets are relatively devoid of the species that drove C/ISON's activity at large $r_\mathrm{H}$ such as CO$_2$ \citep{2013ApJ...776L..20M}.

Using SWAN's observations, we estimate the active area of C/SWAN to be of the order of $0.1~\mathrm{km^2}$ at $r_\mathrm{H}=0.5$~AU based on the vaporization model of \citet{1979M&P....21..155C}. This places a lower limit of 200~m on the nucleus, which is substantially larger than the $\sim 80$~m value constrained by the peak brightness observed by SOHO. A similar size reduction was noted for C/ISON \citep{2014ApJ...782L..37K}, but the reduction for C/SWAN (200~m vs. 80~m) is not as large as C/ISON's ($\sim1.5$~km vs. $\sim100$~m). While the exact numbers are probably not to be trusted, such a distinct difference suggests that C/ISON loses mass more efficiently. The fact that other bright Kreutz comets (e.g. C/Lovejoy) are undetected by SWAN may imply that C/SWAN loses mass more efficiently than its fellows, or that it contains more watery volatile than others.

\subsection{C/2012 U3 (STEREO) and SOHO-2388}

The non-detections by CFHT suggest upper limits of sizes of 240~m and 220~m, respectively for C/STEREO (at $r_\mathrm{H}=1.31$~AU) and SOHO-2388 (at $r_\mathrm{H}=1.34$~AU). The non-detections indicate that C/STEREO and SOHO-2388 have $n\gtrsim7$ during their brightening burst or their brightening bursts start beyond $50~\mathrm{R_\odot}$. Since their light-curves behave similarly to that of C/Lovejoy based on the available data, it is possible that they follow a similar pattern of C/Lovejoy and that their brightening burst stages extend to $\sim 1$~AU.

\begin{figure*}
\includegraphics[width=\textwidth]{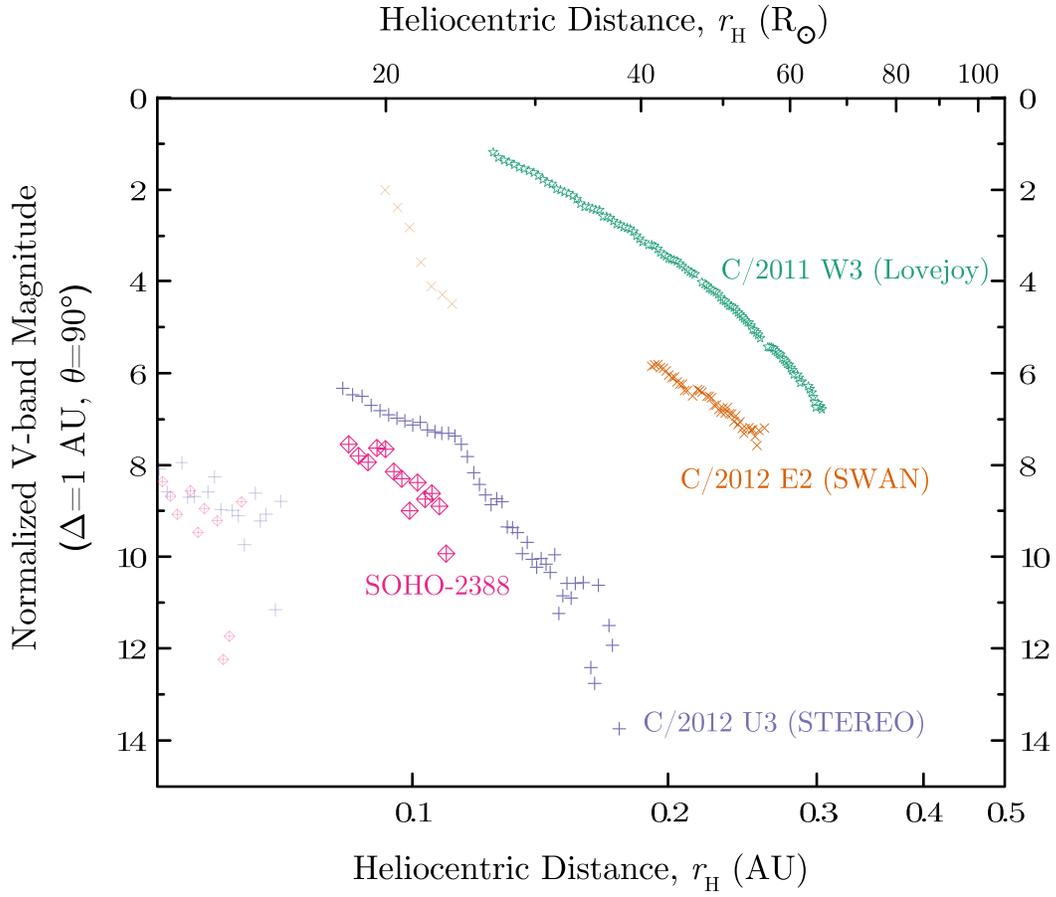}
\caption{Light-curves of C/2011 W3 (Lovejoy), C/2012 E2 (SWAN) and C/2012 U3 (STEREO) derived from HI-1 data. COR-2 light-curves are plotted in lighter colors. There is no COR-2 measurement for C/Lovejoy due to saturation.}
\label{fig-lc-hi1}
\end{figure*}

We also see a discrepancy in brightening rates constrained by different instruments, notably for C/STEREO, where $n$(C3)$=7.9$, $n$(HI-1)$=15.0$ near $25~\mathrm{R_\odot}$, and $n$(C3)$=1.8$, $n$(COR-2)$=4.2$ near $15~\mathrm{R_\odot}$ (Figure~\ref{fig-comet-bandpass} and~\ref{fig-lc-hi1}). Altering $\delta_{90}$ to extreme values (such as $\delta_{90}=100$) produces little change and we conclude that such a discrepancy must be caused by emissions that exist in the bandpass of one instrument but not the other. Following the discussion for C/SWAN, the discrepancy between C3 and HI-1 is most likely due to a blue-ward species such as CN. We can also infer that sodium is probably not the primary driver for the brightening burst, as C3's brightening rate would benefit instead of lag if the emission is originated in the C3 bandpass (such as sodium). The discrepancy between C3 and COR-2 may have a similar nature to C/SWAN's feature at $22~\mathrm{R_\odot}$ and is probably due to erosion of the less-volatile portion of the nucleus. 

\section{Conclusion}

We conducted a search for mini Kreutz comets at moderate heliocentric distance. Assisted by SOHO/STEREO observations, we identified two such comets that should be in our images but are not detected. Combining with the results from previous workers, we concluded that the brightening burst stage of these comets has $n\gtrsim7$ or the brightening burst starts beyond $50~\mathrm{R_\odot}$.

We also compiled a composite set of pre-perihelion light-curves of some bright Kreutz comets that covers to $\sim 1$~AU, including C/1965 S1 (Ikeya-Seki), C/2011 W3 (Lovejoy), C/2012 E2 (SWAN), C/2012 U3 (STEREO) and SOHO-2388. We found at least two distinct shapes: C/Ikeya-Seki and C/SWAN follow $n=4$ and others follow $n\geq7$.

C/SWAN is particularly interesting, as a non-detection at $r_\mathrm{H}=1.06$~AU suggested an outburst ($\Delta m>5$~mag) or a rapid brightening ($n\gtrsim11$) that occur somewhere between $r_\mathrm{H}=1.06$~AU to $0.52$~AU. Although broad-band photometry suggested a steady brightening with $n=4$ to perihelion, a brief surge was observed at $r_\mathrm{H}=0.1$~AU solely in the 650--750~nm range, probably hinting at the erosion of a less-volatile portion of cometary nucleus. We estimated the size of C/SWAN's nucleus to be $<680$~m at $r_\mathrm{H}=1.06$~AU, $\sim 200$~m at $r_\mathrm{H}=0.5$~AU and $\sim 80$~m at $r_\mathrm{H}=0.1$~AU. The reduction of nucleus size from $r_\mathrm{H}=0.5$ to 0.1~AU for C/SWAN is substantially smaller than that of C/2012 S1 (ISON), seemly suggesting that C/SWAN loses mass less efficiently than C/ISON. The fact that other bright Kreutz comets (like C/Lovejoy) were undetected by SWAN may imply that C/SWAN either loses mass more efficiently or contains more water-ice than its compatriots. The early detection of C/ISON at large $r_\mathrm{H}$ and the non-detection of any mini Kreutz comets indicate that the surface layers of mini Kreutz comets are relatively devoid of the species that drove C/ISON's activity at large $r_\mathrm{H}$ such as CO$_2$.

C/STEREO was concurrently observed by COR-2, C3 and HI-1; it brightened more rapidly in HI-1 than C3 at brightening burst, which suggests that a blue-ward species (probably CN) may be the main driver instead of sodium. C/STEREO also brightened faster in COR-2 than C3 at $15~\mathrm{R_\odot}$, which may be similar to C/SWAN's COR-2 surge in nature and is probably due to erosion of the less-volatile portion of the nucleus.

Light-curve analysis is a useful tool to probe the intrinsic properties of comets. This may be especially true for comets with unique orbits like the Kreutz comets, in which the physical/chemical process are accelerated and magnified due to the extreme environment they suffer. In this study we see hints of diversity among the selected comets from light-curves extended to $\sim 1$~AU. Future studies aiming at determining longer segments of pre-perihelion light-curves of a more statistically meaningful set of Kreutz comets may help uncovering the physics behind this diversity.

\acknowledgments

We thank an anonymous referee for his/her very throughout comments that help improve this work. We thank Karl Battams, Matthew Knight and Joe Marcus for helpful discussions, and Jason Jill for maintaining the hardware used for data analyses. We gratefully acknowledge the CFHT operation team and SOHO/STEREO operation teams for their works, as well as all the SOHO/STEREO comet hunters for their contribution. We also thank Eric Christensen and Robert McNaught for helps on the SSS data. Q.-Z. thanks Peter Brown for his encouragement and support, as well as Pauline Barmby and Reto Musci for their helps on the paperwork; he also thanks Summer Xia Han for help fighting his procrastination. Parts of the observational data in this work is based on observations obtained with MegaPrime/MegaCam, a joint project of CFHT and CEA/DAPNIA, at the Canada-France-Hawaii Telescope (CFHT) which is operated by the National Research Council (NRC) of Canada, the Institut National des Science de l'Univers of the Centre National de la Recherche Scientifique (CNRS) of France, and the University of Hawaii. The SSS survey was operated by the Catalina Sky Survey (CSS) in collaboration with the Australian National University. The CSS/SSS surveys are funded by the National Aeronautics and Space Administration under Grant No. NNH12ZDA001N-NEOO, issued through the Science MIssion Directorate's Near Earth Object Observations Program. This work was performed in part with the support of the Natural Sciences and Engineering Research Council of Canada.

%\bibliographystyle{apj}
%\bibliography{man}

\clearpage

\clearpage

\clearpage

\clearpage

\clearpage

\clearpage

\clearpage

\clearpage

\clearpage

\clearpage

\begin{table*}
\begin{center}
\caption{SOHO/STEREO instruments used for Kreutz comet observations, compiled from \citet{2009SoPh..254..387E} and \citet{2010AJ....139..926K}. \label{tbl-spacecrafts}}
\begin{tabular}{lllll}
\hline
Spacecraft & Instrument & Bandpass & Resolution & Field of view \\
\hline
SOHO & Coronagraph 2 (C2) & 400--850 nm & 12''/pix & 2--$6~\mathrm{R_\odot}$ \\
SOHO & Coronagraph 3 (C3) & 400--850 nm & 56''/pix & 4--$30~\mathrm{R_\odot}$ \\
STEREO & Coronagraph 2 (COR-2) & 650--750 nm & 15''/pix & 2--$15~\mathrm{R_\odot}$ \\
STEREO & Heliospheric Imager 1 (HI-1) & 630--730 nm\tablenotemark{a} & 70''/pix\tablenotemark{b} & 3--$23^{\circ}$ along ecliptic \\
\hline
\end{tabular}
\tablenotetext{a}{Some transmission at 370--420~nm and 930--950~nm \citep{2010SoPh..264..433B}.}
\tablenotetext{a}{The pixel size of the detector is 35''/pix, but the observations we used are binned by a factor of two.}
\end{center}
\end{table*}

\clearpage

\begin{table*}
\begin{center}
\caption{Details of the CFHT observations. The $1\sigma$ detection efficiency is determined based on the completeness of detection of artificial moving sources implanted to the actual data (see main text). \label{tbl-obs}}
\begin{tabular}{lllccc}
\hline
Date & RA range & Dec range & Elongation & FWHM & $1\sigma$ efficiency in g' \\
\hline
2012 Sep 19 & 08h40m~to~08h48m & $-10^{\circ}$ to $-7^{\circ}$ & $\sim46^{\circ}$ & $\sim1.0$'' & $21.0$ \\
2012 Sep 20 & 08h36m~to~08h44m & $-10^{\circ}$ to $-7^{\circ}$ & $\sim48^{\circ}$ & $\sim1.5$'' & $21.0$ \\
2012 Sep 22 & 08h36m~to~08h44m & $-11^{\circ}$ to $-8^{\circ}$ & $\sim49^{\circ}$ & $\sim1.1$'' & $22.0$ \\
2012 Oct 16 & 10h11m~to~10h17m & $-20^{\circ}$ to $-14^{\circ}$ & $\sim47^{\circ}$ & $\sim2.0$'' & $22.0$ \\
2012 Oct 17 & 10h11m~to~10h17m & $-22^{\circ}$ to $-16^{\circ}$ & $\sim48^{\circ}$ & $\sim1.7$'' & $22.0$ \\
2012 Oct 18 & 10h11m~to~10h17m & $-22^{\circ}$ to $-16^{\circ}$ & $\sim49^{\circ}$ & $\sim1.4$'' & $22.0$ \\
2012 Oct 20 & 10h19m~to~10h25m & $-22^{\circ}$ to $-16^{\circ}$ & $\sim49^{\circ}$ & $\sim1.4$'' & $21.5$ \\
2012 Oct 21 & 08h19m~to~10h25m & $-23^{\circ}$ to $-17^{\circ}$ & $\sim50^{\circ}$ & $\sim1.2$'' & $22.0$ \\
\hline
\end{tabular}
\end{center}
\end{table*}

\clearpage

\begin{table*}
\begin{center}
\caption{Kreutz comets with STEREO-derived orbits that are searched against the CFHT observations. \label{tbl-adhoc}}
\begin{tabular}{llcl}
\hline
Comet & Data date & Detection limit\tablenotemark{a} & Comments \\
\hline
SOHO-2374 & 2012 Sep 19--22 & - & too far east \\
SOHO-2375 & 2012 Sep 19--22 & - & too far east \\
C/2012 U3 (STEREO) & 2012 Sep 20 & 23.7~g' & within FoV \\
SOHO-2388 & 2012 Sep 22 & 24.0~g' & within FoV \\
SOHO-2404 & 2012 Oct 16--21 & - & $10^{\circ}$ too west \\
\hline
\end{tabular}
\tablenotetext{a}{Computed at SNR=3 at the star-masked combined data.}
\end{center}
\end{table*}

\clearpage

\begin{table*}
\begin{center}
\caption{Orbits of the Kreutz comets that are searched in CFHT observations as derived using STEREO observations. \label{tbl-adhoc-orb}}
\begin{tabular}{lcccccc}
\hline
Designation & $T_p$ (UT) & $q$ & $e$ & $i$ & $\omega$ & $\Omega$ \\
\hline
SOHO-2374 & 2012 Oct 6.86 & 0.0054 & 1.0 & $143.98^{\circ}$ & $81.83^{\circ}$ & $3.12^{\circ}$ \\
SOHO-2375 & 2012 Oct 13.80 & 0.0049 & 1.0 & $144.05^{\circ}$ & $80.55^{\circ}$ & $1.42^{\circ}$ \\
C/2012 U3 (STEREO) & 2012 Oct 31.33 & 0.0053 & 1.0 & $144.14^{\circ}$ & $81.18^{\circ}$ & $1.45^{\circ}$ \\
SOHO-2388 & 2012 Nov 4.84 & 0.0046 & 1.0 & $144.22^{\circ}$ & $80.47^{\circ}$ & $0.32^{\circ}$ \\
SOHO-2404 & 2012 Nov 27.40 & 0.0047 & 1.0 & $144.13^{\circ}$ & $80.55^{\circ}$ & $0.94^{\circ}$ \\
\hline
\end{tabular}
\end{center}
\end{table*}

\end{CJK*}
\end{document}